\documentclass[conference]{IEEEtran}
\usepackage[utf8]{inputenc}

\IEEEoverridecommandlockouts

\usepackage{comment}
\usepackage{environ}     %
\usepackage{etoolbox}    %
\usepackage{graphicx}    %
\usepackage{amsmath,epsfig,epstopdf,amssymb}
\usepackage{adjustbox,lipsum}
\usepackage[hidelinks]{hyperref}
\usepackage{url}
\usepackage{tcolorbox}
\usepackage{adjustbox}
\usepackage{MnSymbol}

\makeatletter   %
\def\@begintheorem#1#2{ \relax
    \topsep 0.3\@IEEEnormalsizeunitybaselineskip plus 0.26\@IEEEnormalsizeunitybaselineskip minus 0.05\@IEEEnormalsizeunitybaselineskip
    \rmfamily\slshape\trivlist%
    \item[]\textsc{\small\noindent #1 #2:} \relax}
\def\@opargbegintheorem#1#2#3{ \relax
    \topsep 0.3\@IEEEnormalsizeunitybaselineskip plus 0.26\@IEEEnormalsizeunitybaselineskip minus 0.05\@IEEEnormalsizeunitybaselineskip
    \rmfamily\slshape\trivlist%
    \item[]\textsc{\small\noindent #1 #2:} \textbf{#3.} \relax}

\def\algorule{\leavevmode\leaders\hrule height .8pt\hfill\kern\z@}
\makeatother

\newtheorem{definition}{Definition}
\newtheorem{example}{Example}
\newtheorem{observation}{Observation}

\usepackage{algpseudocode}
\algnewcommand\algorithmicforeach{\textbf{for each}}
\algdef{S}[FOR]{ForEach}[1]{\algorithmicforeach\ #1\ \algorithmicdo}
\algrenewcommand{\algorithmiccomment}[1]{\hfill \# \begingroup\footnotesize #1\endgroup}

\usepackage{multirow}

\usepackage{balance} %

\usepackage[backend=biber, abbreviate=true, dateabbrev=true, alldates=year, isbn=false, doi=true, urldate=comp, url=false, maxbibnames=9, backref=false, style=numeric-comp, language=american, giveninits=true, sortcites=true, sorting=none]{biblatex}
\AtEveryBibitem{\clearlist{location}} %
\AtEveryBibitem{\clearfield{series}} %
\AtEveryBibitem{\clearlist{language}} %

\usepackage{citation-cleaner}

\usepackage{paralist}

\usepackage[colorinlistoftodos,prependcaption,textsize=scriptsize]{todonotes} %

\usepackage{booktabs,siunitx}

\newcommand{\head}[1]{\par\noindent\textbf{#1:}\space}
\newcommand{\f}[1]{\ensuremath{\text{\emph{#1}}}}

\title{Towards Extending the Range of Bugs That Automated Program Repair Can handle\thanks{This work has been financially supported by the Research Council of Norway through the secureIT project (RCN contract \#288787).}
}

\author{\IEEEauthorblockN{Omar I. Al-Bataineh}
\IEEEauthorblockA{\textit{Simula Research Laboratory} \\
Oslo, Norway \\
omar@simula.no\vspace*{-2ex}}
\and
\IEEEauthorblockN{Leon Moonen}
\IEEEauthorblockA{\textit{Simula Research Laboratory} \\
Oslo, Norway \\
leon.moonen@computer.org}\vspace*{-2ex}}

\makeatletter
\def\ps@IEEEtitlepagestyle{%
  \def\@oddfoot{\mycopyrightnotice}%
  \def\@evenfoot{}%
}
\def\mycopyrightnotice{%
  \hspace*{3mm}\includegraphics[width=2cm]{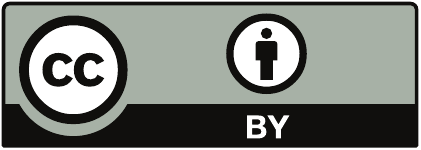}%
  \hspace*{2mm}\raisebox{2.5mm}{%
   	  \parbox{\columnwidth}{\footnotesize This work is licensed under a Creative Commons \\ Attribution 4.0 International (CC BY 4.0) license.}%
   	  \hspace*{-68pt}\mbox{1}\hspace{20pt}\fbox{\parbox{.86\columnwidth}{\footnotesize\textsl{Accepted for publication in the 22nd IEEE International Conference on Software Quality, Reliability and Security (QRS 2022).}}}%
  }%
  \gdef\mycopyrightnotice{}%
}
\makeatother

\begin{document}

\maketitle

\pagestyle{plain}

\noindent\begin{abstract}%
Modern \emph{automated program repair} (APR) is well-tuned to finding and repairing bugs 
that introduce \emph{observable} erroneous behavior to a program. 
However, a significant class of bugs does not lead to such observable behavior
(e.g., liveness/termination bugs, non-functional bugs, and information flow bugs).
Such bugs can generally not be handled with current APR approaches, 
so, as a community, we need to develop complementary techniques. 

To stimulate the systematic study of alternative APR approaches and hybrid APR combinations, 
we devise a novel bug classification system that enables 
methodical analysis of their bug detection power and bug repair capabilities.
To demonstrate the benefits, 
we analyze the repair of termination bugs in sequential and concurrent programs.
The study shows that integrating dynamic APR with formal analysis techniques, 
such as termination provers and software model checkers, 
reduces complexity and improves the overall reliability of these repairs.
\end{abstract}

\begin{IEEEkeywords}%
automated program repair,
bug classification,
non-observable and liveness bugs,
hybrid techniques.
\end{IEEEkeywords}

\section{Introduction}

\noindent
Corrective maintenance, i.e., finding and repairing software defects, is one of the main categories of software maintenance, 
and responsible for a large part of the overall costs of software development~\cite{swanson1976:dimensions}. 
\emph{Automated program repair} (APR) promises to increase developer productivity and drastically reduce the costs of corrective maintenance~\cite{legoues2019:automated, monperrus2018:automatic}.
Despite the advances of APR for real-world programs~\cite{marginean2019:sapfix}, 
these approaches can only handle certain types of bugs 
because they generally rely on dynamic analysis for functional verification, 
where a test suite is used to simulate the input and monitor the output to check correct behavior.
However, this is only viable if the effects of a bug can be observed when executing the program.

Detecting and repairing \emph{non-observable bugs} and \emph{liveness bugs} (i.e., bugs that do not lead to incorrect results or crashes) pose a far greater challenge.
For example, identifying a liveness bug requires finding an infinite execution that will never satisfy the desired liveness property~\cite{alpern1987:recognizing}. 
It is not known how long one would need to run the program to reveal an existing liveness bug,
making it impractical to find such cases using dynamic analysis. 
Another critical challenge that makes detection and repair of liveness bugs notoriously hard 
is that the effects that a liveness bug is triggered are generally unobservable 
(i.e., they typically produce little debugging information).
One option for finding this class of bugs is applying formal program analysis techniques that use correctness specifications to detect liveness bugs. 
Such a rigorous analysis can both help to detect the presence of liveness bugs 
as well as assure the absence of these bugs in automatically generated patches.

The question which (combinations of) techniques will be most effective 
at handling certain bugs is an open research question that forms the foundation of this paper.
To stimulate the systematic study of alternative APR approaches and hybrid APR combinations, 
we devise a novel bug classification system that enables methodical analysis of their
bug detection power and bug repair capabilities.
In earlier work, various bug classification schemes were developed to understand when and why specific bugs arise, and how they are fixed. 
These classifications use a number of criteria, such as cause-impact~\cite{li2006:have,tan2014:bug}, 
severity-priority~\cite{serrano2005:bugzilla}, 
and bug complexity~\cite{cotroneo2016:how}.
However, since they were designed for different goals, they do not capture the properties required to determine if a bug is amenable to a particular technique. 
To that end, we introduce a bug classification system explicitly aimed at comparing different techniques and evaluating the feasibility of their integration.

\head{Contributions}
(1) We propose a \emph{novel bug classification system} based on three fundamental properties: 
\emph{bug observability}, \emph{bug reproducibility}, and \emph{bug tractability}.
This classification provides the APR community with a tool to 
methodologically explore and compare alternative and hybrid APR approaches by
(i) analyzing the detection power of different bug detection techniques, 
(ii) distinguishing APR approaches based on their bug repair capabilities, 
and (iii) providing a common terminology that helps identify gaps in current APR research.

\noindent (2) We discuss four \emph{APR approaches} that can handle different classes of bugs: \emph{dynamic APR}, 
\emph{static APR}, \emph{dynamic-static APR}, and \emph{formal APR}. 
Moreover, we identify the conditions under which each approach can be effectively applied. 

\noindent (3) To demonstrate the benefits of our method, 
we study \emph{termination bugs} in sequential and concurrent programs, 
and sketch novel \emph{hybrid APR algorithms} for repairing such bugs.
The study shows that termination bugs in \emph{sequential} programs 
can be effectively addressed using \emph{dynamic-static APR},
by first generating plausible patches using test cases and then using termination provers~\cite{giesl2014:proving,chen2015:synthesising,brockschmidt2016:t2} to check their correctness.
The non-deterministic nature of termination bugs in \emph{concurrent} programs makes them 
challenging for dynamic analysis, and they are best addressed with \emph{formal APR}
that combines termination provers with software model checkers~\cite{jhala2009:software,
godefroid1997:verisoft,holzmann1997:model,havelund2000:model,musuvathi2002:cmc,thompson2010:software,baranova2017:model}.

\section{Bug Classification Schemes}

\noindent
There are many different ways to expose bugs in programs, including manual inspection, dynamic analysis (testing), static analysis, 
model checking, or a combination of these techniques.
Effective bug classification schemes can help understand why bugs arise and how to fix them.
Classification can also help identify the most appropriate analysis technique for handling each class of bugs.
Next, we discuss three existing bug classification systems, analyze their limitations, and introduce a new classification system that addresses them:
\begin{compactenum}
\item \emph{Cause-impact criteria}~\cite{li2006:have,tan2014:bug}: 
Bugs are classified based on their cause: algorithmic, concurrency, memory, generic programming, and unknown, as well as based on their impact: security, performance, failure, and unknown.

\item \emph{Severity and priority criteria}: 
This classification is used in many bug tracking systems~\cite{serrano2005:bugzilla}.
Severity indicates the impact of the bug on the program's functionality and can be categorized as critical, major, moderate, minor, etc.
Priority indicates how soon the bug should be fixed and is categorized into levels such as low, medium, and high.

\item \emph{Bug complexity criteria}~\cite{cotroneo2016:how}:
These criteria distinguish four main categories: 
(i) easy to detect, easy to repair bugs, 
(ii) easy to detect, difficult to repair bugs,
(iii) difficult to detect, easy to repair bugs, and
(iv) difficult to detect, difficult to repair bugs.

\end{compactenum}
These three existing bug classification systems were not designed for comparing the capabilities and limitations of different bug detection or program repair techniques. 
As a result, their criteria do not capture the specific properties needed to determine whether a bug $b$ is amenable to a particular technique $T$. 
To address this gap, we propose a new bug classification system that is based on three key properties of bugs, 
namely \emph{bug observability}, \emph{bug reproduciblity}, and \emph{bug tractability}.
In Sections~\ref{sec:detection} and~\ref{sec:APR}, these properties are then used to analyze the power of different bug detection techniques and APR approaches.

Before proceeding further, let us first define a program bug.
We base ourselves on a specification of \emph{expected behavior}: the expected responses (output) of the program to a given input.

\begin{definition}[Program bug]
Let $P$ be a program, $I$ a set of inputs, and $\varphi_{beh}$ be a specification of expected behavior of $P$. 
We say that $P$ suffers from a bug %
iff there exists at least one input $i \in I$ that leads to an execution trace %
under which program $P$ violates $\varphi_{beh}$, formalised as $(P, i) \not\models \varphi_{beh}$.
\end{definition}

Since a complete specification of a program's expected behavior is often not available,  
test cases are generally used to model the expected behavior of a program $P$. 
We assume there  exists $(i, o_{exp})$, where $o_{exp}$ is the expected output for input $i \in I$. 
When the observed output $o_{obs} = (P, i)$ does not match the expected output $o_{exp}$, we say that $P$ contains a bug.

\begin{definition}[Observable bug]\label{observability}
Let $P$ be a program containing bug $b$.
We say that $b$ is an \emph{observable bug} iff 
there exists an execution of $P$ where, in a finite number of execution steps, the erroneous behavior of $b$ can be seen by an observer $O$. 
\end{definition}

\begin{definition}[Classifying bugs by observability]
We classify bugs based on the notion of observability in three types:
\begin{compactenum}
\item \emph{observable bugs} whose erroneous behavior is fully observable in finite execution steps (e.g., arithmetic bugs),
\item \emph{partially observable bugs} whose erroneous behavior is only partially observable at runtime because the faulty trace is infinite, so not all output of the program can be observed  (e.g., termination bugs), 
\item \emph{non-observable bugs} whose erroneous behavior is fully unobservable at runtime (e.g., non-functional bugs).
\end{compactenum}
\end{definition}

Note that observability is a relative notion that depends on the observation power of $O$.
In the simplest case, the observer can witness the output produced by $P$ and the corresponding execution time. 
If we increase the observation power of $O$ (i.e., the amount of information $O$ can gather about the program's execution), 
some non-observable bugs may become observable.
For example, bugs that adversely affect the memory or energy consumption can easily go unnoticed during the execution of the program.
Such bugs can be exposed by using monitoring at the virtual machine or operating system level~\cite{hebbal2015:virtual,dovgalyuk2017:qemubased,gregg2020:systems,gregg2019:bpf}.
Alternatively, the program can be augmented with additional variables and checks that help to keep track of these non-functional aspects at runtime~\cite{al-bataineh2021:monitoring}.
However, increasing observation power also increases the chance of affecting the program's execution~\cite{mytkowicz2008:observer}.

We distinguish five common types of observable erroneous behavior:
$EB = \{$\f{crash}, \f{exception}, \f{incorrectResult}, \f{softHang}, \f{hardHang}$\}$.
While most types in $EB$ are easy to understand, 
we will define the notions of \emph{soft} and \emph{hard} hang bugs.
Hang bugs are a particular type of bugs that concern (temporary or permanent) lack of progress in observable behavior~\cite{wang2008:hang,dean2015:automatic}.
Hang bugs can have various causes, such as iteration errors or communication deadlocks.
To define hang bugs, we use a temporal specification~\cite{pnueli1977:temporal,manna1992:temporal} that checks if any of the locations where the program might terminate can be reached.

\begin{definition}[Halting Statements]\label{HltStatements} 
We refer to a statement $s$ in a program $P$ as a \emph{halting statement} iff the expected behavior of $P$ is that execution terminates after executing statement $s$.
\end{definition}
Examples of halting statements include special termination statements such as \texttt{\small exit}, or simply the final statement in a program.
Observe that programs whose expected behavior is to never terminate have no halting statements (e.g., a webserver).

\begin{definition}[Hang bugs]\label{HangBugs}
Let $P$ be a program with a set of inputs $I$ and $H$ be the set of halting statements of $P$. 
Let $\varphi_{temp}$ be a temporal property that puts an upper bound on the execution time of $P$, 
and $\varphi_{reach}$ be a temporal property that checks whether $P$ reaches a halting statement.
Let also $EB' = EB ~ \backslash ~ \{\f{softHang}, \f{hardHang}\}$. 
We distinguish:
\begin{compactenum}
\item \textbf{Soft hang bugs} occur when there exists an input $i \in I$ that makes $P$ unresponsive for a finite amount of time before execution is resumed and a halting statement is reached:
\vspace{-1.3ex}$$  %
S: (P, i)  \not \models \varphi_{temp} ~\land  (P, i) \models \varphi_{reach} \land ~\f{output(P, i)} \not\subseteq EB'
\vspace{-1.3ex}$$  %
\item \textbf{Hard hang bugs}, also known as \textbf{termination bugs}, occur when there exists an input $i \in I$ that makes $P$ unresponsive for an unbounded amount of time, never resuming to normal execution or reaching a halting statement:
\vspace{-1.3ex}$$  %
H: (P, i)  \not \models \varphi_{temp} ~\land  (P, i) \not\models \varphi_{reach} ~ \land  ~ \f{output(P, i)} \not\subseteq EB'
\vspace{-.5ex}$$  %
\end{compactenum}
\end{definition}
Hang bugs are also referred to as \emph{liveness violations} in model checking and formal program analysis literature~\cite{lamport1977:proving,alpern1985:defining,killian2007:life,li2010:tcheck}, and we will use these terms interchangeably in the remainder.

We now turn to discuss the property of bug reproducibility.

\begin{definition}[Bug reproducibility]
Let $P$ be a program containing a bug $b$ and $t_b$ be a test case that exposes $b$.
We say that $b$ is a \emph{reproducible} or \emph{deterministic bug} 
iff every time program $P$ is executed under test $t_b$, bug $b$ is exposed and the same erroneous behavior is observed. 
On the other hand, we say that $b$ is a \emph{hard-to-reproduce} or \emph{non-deterministic bug} 
iff bug $b$ is exposed in rare circumstances when repeating the execution of $P$ under test $t_b$ 
(i.e., the result of program $P$ depends not only on the code of $P$ but also on the timing of the execution).
\end{definition}

\begin{table*}
\small\centering
\caption{A summary of the key properties for bug classification with their attributes and impact} \label{table:propertiesArritbutes}
\vspace*{-2ex}%
\begin{tabular}{cccc} 
\toprule
\textbf{bug property} & \textbf{property attributes} & \textbf{impact on bug detection techniques}  \\
\midrule
observability   & $\{$observable, partially-observable, non-observable$\}$ & affects detection power  \\
reproducibility & $\{$easy-to-reproduce, hard-to-reproduce$\}$ & affects efficiency and scalability  \\
tractability    & $\{$shallow, deep, liveness$\}$ & affects detection power and efficiency  \\
\bottomrule
\end{tabular}%
\vspace*{-2ex}%
\end{table*}

Reproducible bugs are easy to detect, provided that the bug is observable (see Definition \ref{observability}). 
Not surprisingly, hard-to-reproduce bugs are also hard to detect.
Arithmetic bugs are examples of \emph{easy-to-reproduce} bugs, while concurrency bugs are examples of \emph{hard-to-reproduce} bugs.
The last property we study is \emph{bug tractability}, which depends on the \emph{depth} of the bug and the size of the faulty trace it produces. 
\begin{definition}[Bug tractability]\label{TractDef}
Let $P$ be a program containing bug $b$ and $L$ be the set of reachable locations of $P$ and $\ell_{b} \in L$ be the buggy location to the bug $b$.
We say that the trace of $b$ is a \emph{tractable trace} iff for each execution of $P$ that is buggy to $b$, the number of execution steps that are required to reach $\ell_{b}$ is bounded
and that $\ell_{b}$ is not part of a loop that can be executed infinitely often.
On the other hand, we say that the trace of $b$ is \emph{intractable} iff  $\ell_{b}$ is visited infinitely often during the execution of $P$ (i.e., $\ell_{b}$ is part of an infinite loop).
\end{definition}

The presence of loops plays a crucial role in determining the tractability of a bug.
The size of faulty traces for non-loop programs is typically shorter than those in loop programs.
Based on the size of the faulty trace, we can further distinguish the class of tractable bugs: 
(i) \emph{shallow bugs} are tractable bugs with finite short faulty traces, 
(ii) \emph{deep bugs} are tractable bugs with finite but long faulty traces.
For example, a bug in a loop program $P$ that does not occur until a vast number of iterations are executed can be viewed as an example of a deep bug.

\head{Bug Classification System} 
Table \ref{table:propertiesArritbutes} summarises the three properties in our classification system with their distinguishing attributes and impact on bug detection. 
Our classification system associates each bug with a three-tuple of concrete attributes for 
\emph{$\{$observability, reproducibility, tractability$\}$}.
For example, an arithmetic bug has the properties: 
\emph{$\{$observable, easy-to-reproduce, shallow$\}$}.

\section{\label{sec:detection}Bug Detection Techniques}

\noindent
Various bug detection techniques can be used to expose bugs in programs. 
In this work, we are interested in studying three well-known bug detection techniques: dynamic analysis, static analysis, and model checking.
While dynamic analysis detects bugs in programs by executing them, static analysis and model checking use different techniques in bug detection that perform bug checking statically, without running the program.
We start by discussing the requirements needed to expose bugs in each technique, the advantages and disadvantages of the techniques, and the theoretical foundation and detection power of each (i.e., the classes of bugs that each technique can handle).

\subsection{Dynamic Analysis}

\noindent
Dynamic analysis is a technique to identify bugs and vulnerabilities in programs by exercising various runs through the program based on valid inputs. 
Dynamic analysis can be performed using a test suite, which can be developed manually or via test case generation, or through fuzzing, which systematically explores a large amount of automatically generated tests.   
Fuzzing is one of the most common methods used to find vulnerabilities in programs~\cite{miller1990:empirical, miller2006:empirical}.
Early fuzz testing was based on sending random inputs to a program to check if it could be made to crash. 
The techniques have evolved to systematically explore the input space using knowledge from the source code or input formats to discover bugs that are hidden deep in the code. 

Dynamic analysis has several advantages over the other program analysis techniques: 
(i) the program behavior can be monitored, and bugs can be exposed while the program is running; 
(ii) it allows for analysis of programs for which we do not have access to the actual code; 
(iii) it can be conducted against any program; and last but not least, 
(iv) it can identify bugs that are hard to find using static analysis. 
We now discuss the conditions under which bugs may be discovered in dynamic analysis.

\begin{definition}[Bug detection in dynamic analysis]\label{detectability}
Let $P$ be a program containing bug $b$, and $D$ be a dynamic program analysis checker (i.e., an automated testing tool such as a fuzzer).
We say that $b$ is detectable in $D$ iff 
\begin{compactenum}
\item $P$ is given in an executable form,
\item bug $b$ is \emph{observable} in some executions of $P$, 
\item there exists a test suite $T$ containing at least one failing test $t$ 
by which %
bug $b$ can be exposed in $P$.
\end{compactenum}
\end{definition}

However, many bugs whose detection requires the analysis of infinite traces or the satisfaction of complex composite properties cannot be found using an approach solely relying on test cases.
Examples of such classes of bugs include 
(i) bugs in non-executable programs, 
(ii) liveness bugs such as termination and starvation bugs, and 
(iii) non-observable bugs such as non-functional and information flow bugs~\cite{sabelfeld2003:languagebased,smith2007:principles}.

\begin{observation}
Dynamic analysis techniques can handle observable classes of bugs with finite execution traces, provided that a testing mechanism is implemented by which the bug can be exposed, and provided that the program is executable.
\end{observation}

To address these limitations, we need to pair dynamic analysis with complementary bug detection techniques to improve the detection power of the approach.
The remainder of this section discusses static analysis and model checking, which instead of using the program itself as in dynamic analysis, analyzes abstractions of the program, to improve observability and tractability and enable the detection of deeper bugs~\cite{david2016:danger}. 

\subsection{Static Program Analysis}

\noindent
Static program analysis is an approach for analyzing a computer program without actually executing it.
The most significant advantage of static analysis is the ability to quickly and automatically examine the complete code of the program to find flaws that might be missed by dynamic analysis.
The literature on static program analysis for bug detection is rich and mature~\cite{dsilva2008:survey,bessey2010:few,sadowski2018:lessons}. 
Many of these techniques build on automatically evaluated analysis rules and bug detection patterns that capture the general conditions under which specific bugs can occur, providing a systematic way for their detection.

To capture the notion of bug detectability in static analysis, one needs to ensure the availability of the source code of the buggy program at which the bug $b$ occurs and the availability of some valid solid theory for the detection of bug $b$.

\begin{definition}[Bug detection in static analysis]
Let $P$ be a program containing bug $b$ and $S$ be a static program analyzer that can be used to expose bugs of type $b$.
We say that $b$ is a bug detectable by analyzer $S$ if all of the following conditions hold:
\begin{compactenum}
\item the source code of $P$ is available, and
\item $S$ contains sound and complete detection method for $b$, and
\item $P$ is written in a language that is accepted by $S$. 
\end{compactenum}
\end{definition}

The increasing complexity of (loop) programs and the large variety of vulnerabilities make it difficult for static code analyzers to detect and identify vulnerabilities in a precise manner.
One of the most significant disadvantages of the static code analysis methodology is the presence of false-positive warnings: the tool may signal possible bugs where there are none.
However, reducing the number of false positives in static analysis tools is still an open problem. 
\begin{observation}
Static checkers can handle observable, non-observable, reproducible, and non-reproducible bugs, 
provided that the checker is built based on some solid mathematical foundation,
and provided that the source code of the program is available in a programming language accepted by the checker.
\end{observation}

Dynamic and static analysis techniques have different bug detection powers as they rely on distinct assumptions and use orthogonal detection methods.
Overall, these techniques have complementary strengths and weaknesses that are worth combining to improve the reliability of APR systems.

\subsection{Model Checking}

\noindent
Model checking~\cite{burch1992:symbolic, berard2001:systems, clarke2018model} is an automated formal method for checking whether a finite-state model of a system meets a given specification.
The technique has been used successfully to debug complex computer hardware, concurrent systems, and real-world safety-critical systems.

Model checking has several advantages. 
It can detect errors that would be very difficult to notice with other methods, such as in concurrent programs. 
The properties that can be verified are more expressive than with traditional testing, depending on the formalism used to express them.
For example, properties that require something to happen infinitely often, or properties that require that some alternative is always available. 
In addition, because every possible behavior of the model is checked, the result is inevitable, provided that the model checking tool itself has no serious errors.

Model checking can be an expensive procedure in a repair process because of its exhaustive nature.
Expressing both the model of the system and the properties formally requires great care and expertise. 
Moreover, one of the most significant problems with model checking in practice is the so-called ``state explosion problem'': When the number of state variables in the system increases, the size of the system state space grows exponentially.
However, abstractions can be applied to bring the verification within feasible bounds of model checking technology.

Software model checking tools~\cite{jhala2009:software,godefroid1997:verisoft,holzmann1997:model, havelund2000:model,musuvathi2002:cmc,thompson2010:software,baranova2017:model} verify the correctness of software models in a rigorous and automated fashion.
Most tools construct a (symbolic) reachability graph for the program-under-analysis 
(i.e., a graph that contains reachable run-time states of the program) \emph{without} running the program. 
This graph is then used to check if a property of interest holds. 
They typically implement sophisticated data structures that enable clever search algorithms and optimizations.

The answer returned by a model checker is either a notion of a successful verification (i.e., the specification holds), 
or a counterexample – an execution path that violates a given property.
However, if the program being verified has an infinite state space, certain types of abstractions are needed, or the analysis may simply not
terminate. %

\begin{definition}[Bug detection in model checking]\label{ModelcheckingDef}
Let $P$ be a program containing bug $b$ and $\f{SMC}$ be a software model checker that can be used to expose bugs of type $b$. 
We say that $b$ is a detectable bug in the model checker $\f{SMC}$ iff:
\begin{compactenum}
\item a formal property $\varphi_\f{b}$ is available that is written in the input specification language of $\f{SMC}$, which captures the conditions under which $b$ can occur, and

\item the source code of $p$ is available, and

\item $P$ has finite states or an equivalent finite abstract program $P_{abs}$ can be constructed for the properties of interest, and

\item the program $P$ or its reduced equivalent program $P_{abs}$ is written in a modeling language that is acceptable by $\f{SMC}$.

\end{compactenum}
\end{definition}

\begin{table*}
\small\centering
\caption{A summary of key differences and similarities of dynamic analysis, static analysis, and model checking} \label{table:comparison}
\vspace*{-2ex}%
\resizebox{\linewidth}{!}{%
\begin{tabular}{cccc} 
\toprule
\textbf{criteria} & \textbf{dynamic analysis} & \textbf{static analysis} & \textbf{model checking} \\
\midrule
bug detection mechanism & test cases & pattern-based specifications & formal correctness specifications \\ 
accuracy of the analysis & accurate & inaccurate (suffers from false positives) & accurate relative to the accuracy of model\\
the need of code availability & not needed & needed to perform the analysis & needed to construct the model\\
the need of bug observability & needed to detect the bug& not needed (performs code analysis) & not needed (performs state analysis) \\
code coverage of the program & incomplete code coverage & complete code coverage & complete code coverage\\
the need of code executability & needed to detect the bug& the program will not be executed & the program will not be executed \\
language dependency & language-independent & language dependent & language dependent \\
automation of the analysis & can be automated (fuzz testing) & fully automated & a model needs to be manually written \\
\bottomrule
\end{tabular}}%
\vspace*{-2ex}%
\end{table*}

\begin{observation}
Model checking tools can handle observable, non-observable, reproducible, and non-reproducible bugs, provided that the size of the program is finite or a sound abstraction can be developed to bring the program within the feasibility bound of model checking, and provided that a specification is available for the bug of interest.
\end{observation}

\head{Bug Detection Properties}
Based on the characteristics of the three bug detection techniques discussed, one can make the following general observations. 
Model checking is more expensive than static analysis, requiring longer running times and more resources. 
Static analysis is not as accurate as model checking, 
and testing is not as complete as model checking. %
Testing suffers from coverage challenges: it is challenging to cover all possible executions of the program, 
in particular for programs with an infinite input space where this becomes prohibitively expensive.
To ensure the correctness of the program for all inputs, a correctness specification must exist that can be formally analyzed.

Table \ref{table:comparison} summarises the key differences and similarities between the three program analysis techniques using several criteria: code coverage, the need for code executability, accuracy of the analysis, the need for bug observability, the need for code availability, and automation of the technique.
By code coverage, we mean the number of feasible execution paths of the program that the technique can cover during the analysis, and accuracy indicates whether the detected bug is a real bug.

\section{\label{sec:APR}APR Approaches}

\noindent
This section describes four APR approaches that can handle different classes of bugs. 
The four approaches combine the detection power of dynamic analysis, static analysis, and model checking techniques to improve the reliability of existing APR techniques.
We discuss the applicability of these approaches to three classes of bugs:  arithmetic bugs (observable bugs), non-functional bugs (non-observable bugs), and liveness bugs (partially observable bugs).

An APR approach generally consists of four steps: fault identification, fault localization, 
patch generation, and patch validation.
The most challenging  step in the APR process is the patch validation step, 
in which the generated patch is extensively evaluated to ensure that the bug is resolved, and
that the patch does not introduce any unwanted behavior.
In dynamic APR, as the name implies, the patch validation step is primarily performed using test cases. 
Since these rarely capture the expected behavior in full detail, 
the technique suffers from the so-called patch overfitting problem, 
where the patched program may pass the tests in the given test suite, 
while it is failing for valid inputs not covered by the test suite.
It is therefore desirable to combine the power of different analysis techniques
while taking into account the distinctive properties of each class of bugs.
This leads to our examination of the following four APR approaches:
\begin{compactenum}
\item \emph{dynamic APR}: in which fault identification and patch validation are performed 
using dynamic analysis techniques. GenProg~\cite{legoues2012:genprog} is an example of a dynamic APR tool;

\item \emph{static APR}: in which fault identification and patch validation are performed 
using static analysis techniques;

\item \emph{dynamic-static APR}: in which fault identification and patch validation are performed 
using a combination of static and dynamic analysis, i.e., using test cases and static analysis tools developed for the same class of bugs; 

\item \emph{formal APR}: in which fault identification and patch validation are performed 
using formal methods and verification techniques such as model checking.

\end{compactenum}
A hybrid APR approach aims to improve the overall quality of the generated repairs and alleviate the patch overfitting challenge of dynamic APR systems.

\subsection{Arithmetic Bugs}

\noindent
Arithmetic calculations affect a wide variety of applications, including safety-critical systems such as control
systems for vehicles, medical equipment, and industrial plants.
The key properties of arithmetic bugs can be summarised as follows:
\begin{compactenum}
\item Arithmetic bugs are observable classes of bugs or can be easily made observable to external observers: the root causes of arithmetic bugs are limited and easy to identify.
\item Arithmetic bugs are tractable bugs with finite traces.
\item Arithmetic bugs introduce various erroneous behavior to the program in which they occur: they may cause the program to crash or may produce incorrect outputs.
\end{compactenum}
These properties make them directly amenable to dynamic (i.e., test-based) APR.
Note that many of the available dynamic APR systems rely explicitly or implicitly on observability strategies to expose this class of bugs (e.g., integer overflow, division by zero, etc.). 
There are also several static analysis tools that can handle arithmetic bugs.
Thus, arithmetic bugs can be repaired using dynamic APR, static APR, or dynamic-static APR. 
These repair approaches differ mainly in the correctness specification used to validate generated patches for the detected arithmetic bug.
In dynamic APR, the specification is captured by the test cases, while in static APR, the specification is captured by the formal bug detection rules.

\begin{example}%
\textbf{Integer overflow (IO)} is a type of arithmetic bug that occurs when the computation of an arithmetic operation, such as multiplication or addition, exceeds
the maximum size of the integer type used to store it.
IO bugs are an observable class of bugs, and thus they are amenable to dynamic APR. 
IO bugs are also amenable to static APR, and there are several reliable static analysis tools available that can address IO bugs~\cite{muntean2021:intrepair,al-bataineh2021:reliable}.
Thus, devising a hybrid static-dynamic APR system for IO bugs is feasible and will help increase confidence about the soundness of the generated repairs.
\end{example}

\begin{observation}
Arithmetic bugs are observable, tractable, and reproducible classes of bugs with finite execution traces. 
They are amenable to both dynamic and static APR since the root causes of arithmetic bugs are easy to identify.
\end{observation}

\subsection{Non-functional Bugs}

\noindent
Non-functional bugs~\cite{jin2012:understanding,radu2019:dataset,al-bataineh2021:monitoring} are a class of bugs that affect the way a program operates, 
rather than the functional behavior of the program.
Inefficiently written loops in programs and synchronization issues in concurrent programs 
(i.e., using a large unnecessary number of locks) can be viewed as examples of non-functional bugs.
Non-functional bugs are as important as functional bugs.
For example, energy consumption saving is getting more urgent, 
particularly for applications running on embedded systems and IoT in Smart Cities.

Fixing non-functional bugs is generally more complex than fixing functional bugs, since non-functional bugs can hide themselves well in the code. 
While most functional bugs can be detected through observing the erroneous behavior of bugs, a large percentage of non-functional bugs are detected through manual code review~\cite{fagan1976:design,gilb1993:software,bacchelli2013:expectations}. 
Non-functional bugs usually do not generate incorrect results or crashes. 
Therefore, they cannot be observed by checking the program output. 
The key properties of non-functional bugs can be described as follows.
\begin{compactenum}
\item Non-functional bugs are generally non-observable classes of bugs: they do not introduce direct observable erroneous behavior to the program in which they occur.
\item Non-functional bugs increase the anticipated running cost of a program (execution time, memory and energy consumption, etc.) due to the inefficient use of resources. 
\end{compactenum}
Depending on what quality attributes are considered, 
programs may suffer from many different types of non-functional bugs. 
For example, consider the class of non-functional bugs that adversely affect the run-time costs of executing the program, such as execution time, 
memory consumption, and energy consumption.
To expose such types of bugs, 
the program may need to be augmented with additional variables or 
online monitors that can be used to observe aspects at runtime~\cite{al-bataineh2021:monitoring}.

Therefore, there is a need to develop effective bug detection tools that can be used to expose non-functional bugs at the early stages of the software development life cycle.
Specifically, this requires efficient profiling techniques and oracles that help decide whether the program's non-functional requirements are met under a particular workload. 
Unfortunately, the lack of effective test oracles for non-functional bugs is a well-known problem that will need to be addressed in the future.

\begin{observation}
Non-functional bugs are an example of non-observable classes of bugs 
(i.e., a program that suffers from a non-functional bug executes normally and terminates normally). 
Thus, they are not directly amenable to traditional dynamic bug detection techniques that rely on test cases and observing a program's outputs.
\end{observation}

\subsection{Liveness Bugs}\label{sec:LivenessBugs}

\noindent
In this section, we discuss a class of bugs that has received little attention from the APR community, 
namely liveness bugs.  
A \emph{liveness property} asserts that ``something good will eventually occur when executing a program''~\cite{lamport1977:proving,alpern1985:defining}. 
Freedom of starvation and program termination are examples of liveness properties.
A program that violates a liveness property cannot make progress and thus suffers from a \emph{liveness bug}. 

Two fundamental properties make detecting and repairing liveness bugs far more challenging than other classes of bugs. 
First, the behavioral effects of triggering a liveness bug are generally unobservable.
Second, identifying a liveness bug requires finding an infinite execution that will never satisfy the desired liveness property~\cite{alpern1987:recognizing}, 
making it impractical to find such bugs using dynamic analysis.
Therefore, detecting and repairing liveness bugs generally require more sophisticated repair algorithms since they must be able to generate a finite representation of infinite counterexamples.

To better understand the complexity of repairing liveness bugs, we study a subset of liveness bugs known as termination bugs. 
There are two advantages to this choice:
On the one hand, a termination bug is a specific type of liveness bug whose repair is essential for ensuring software reliability. 
On the other hand, by examining techniques for handling termination bugs, we gain knowledge that can help address other liveness bugs. 
Termination bugs have the following specific properties:
\begin{compactenum}
\item Termination bugs are \emph{partially observable}: 
    an observer monitoring the behavior of a non-terminating loop program will not witness any erroneous behavior 
    but rather experience unexpectedly long execution times. 
    The only observable behavior of termination bugs is that the program becomes \emph{non-responsive at runtime}.
\item Termination bugs have \emph{infinite faulty traces}: a counterexample to a termination property violation is infinite.
\end{compactenum}

\begin{observation}
Termination bugs are an example of partially observable bugs with infinite execution traces.
Termination bugs are amenable to dynamic-static APR and static APR.
\end{observation}

\section{Hybrid APR for Termination Bugs} \label{sec:HybridTermination}

\noindent
This section describes a hybrid program repair approach for termination bugs that combines the strengths of termination provers with those of software model checkers.
Such a combination has two key advantages. 
First, it considerably reduces the overall computational complexity of the problem by avoiding the exhaustive exploration of the program's input space. 
Second, it helps avoid the known overfitting problem by generating verified repairs for termination bugs.

The presence of loops in programs can complicate the detection of certain classes of bugs. 
Recall the two types of hang bugs from Definition~\ref{HangBugs}; 
It is not clear how one can distinguish between the following two types of loops using a dynamic analysis technique: 
(i) inefficiently written loops that introduce a soft hang bug, and
(ii) incorrect infinite loops that introduce a hard hang or termination bug.

Earlier work~\cite{legoues2015:manybugs} that evaluated the effectiveness of different APR tools on the ManyBugs and IntroClass datasets, 
used a simple timeout mechanism to handle termination bugs in these two datasets:
when the execution time of a program exceeds some pre-specified period, 
they consider the program to be \emph{likely} non-terminating due to an infinite loop.
Marcote and Monperrus instrument loops with iteration counters that are monitored to detect infinite loops~\cite{marcote2015:automatic}.
Both options \emph{can} lead to false conclusions about the program under analysis 
(i.e., even when the watchdog triggers, the program may not have a termination bug 
but suffer from a soft hang bug).
Moreover, hang bugs can have complicated causes: 
programs that become unresponsive may contain deadlocks, infinite loops, 
or other bugs that lead to non-termination but are \emph{not} infinite loops.

An effective solution for addressing termination bugs is to apply \emph{termination provers}: 
tools that can check combinations of many complex termination criteria.
They take a program as input and return one of three answers: 
\emph{terminating} ($\f{TR}$), \emph{non-terminating} ($\f{NT}$), or \emph{unknown} ($\f{UN}$).
In general, when the prover returns definite answer for a given program (i.e., $\f{answer} \in \{\f{TR}, \f{NT}$\}), 
the answer is valid with high confidence. 
Termination provers have been successfully used to analyse termination of a wide variety of loop programs~\cite{berdine2007:variance, chawdhary2008:ranking,tsitovich2011:loop, gulwani2009:controlflow, gulwani2009:speed, bradley2005:linear, cousot2005:proving, gupta2008:proving, harris2010:alternation, kroening2010:termination, podelski2004:transition}.

\begin{definition}[Valid Termination Bug Repair]
Let $P$ be a buggy non-terminating program with a set of inputs $I$,
$\varphi_{beh}$ a specification of expected behavior of $P$, 
and $\varphi_{reach}$ a specification that checks the reachability of some halting statement of $P$. 
We say that $P'$ is a valid repair of $P$ iff for every input $i \in I$ we have $(P', i) \models \varphi_{beh}$ and $(P', i) \models \varphi_{reach}$.
\end{definition}
In other words, the patched version $P'$ should preserve the expected behavior of $P$, and it should terminate.

\subsection{Termination Bugs in Sequential Programs}\label{termSeq}

\noindent
To repair termination bugs in sequential programs, we first generate plausible patches,
and then use termination provers to check the correctness of these patches. 
AProVE~\cite{giesl2014:proving} and 2LS~\cite{chen2015:synthesising} are among the most reliable candidates to analyze the termination of sequential programs. 

\begin{definition}[Validity of Patches for Termination Bugs in Sequential Programs] \label{ValidityDef} 
Let $P_s$ be a non-terminating sequential program and $T = (T_p \cup T_f)$ be a test-suite consisting of passing test cases $T_p$ and failing test cases $T_f$.
Let $P'_s$ be a candidate patch of $P_s$ and $\f{TP}$ be a termination prover that returns one of the verification answers $\{\f{TR}, \f{NT}, \f{UN}\}$. %
We say that $P'_s$ is a valid patch of $P_s$ iff all of the following conditions hold:
\begin{compactenum}
\item all failing test cases from $T_f$ pass on program $P'_s$, 
\item none of the passing test cases from $T_p$ fail on program $P'_s$, 
\item termination prover $\f{TP}$ returns ``TR'' when analyzing termination of $P'_s$. 
\end{compactenum}
\end{definition}
The soundness of generated patches for termination bugs in sequential programs can be captured formally as:
\begin{equation}\label{SpecSeq}
\varphi_\f{seq} = 
(\forall_{t \in T} (P'_s \vdash t) \land \f{TP}(P'_s) = \f{TR})
\end{equation}
where $T$ is the set of available test cases, and 
$P'_s \vdash t$ indicates that patch $P'_s$ runs successfully against test $t$.

Figure~\ref{alg:RepairAlgSeq}  sketches a hybrid repair procedure for termination bugs in sequential programs.
The algorithm takes five inputs: the buggy sequential program $P_s$, 
a specification of expected behavior $\varphi_\f{beh}$, 
a test suite $T_s$, a termination prover \f{TP}, 
and the allocated time budget $\f{TimeBudget}$. 
It uses two functions: 
(i) $\f{faultLocalizer}(P_s, \f{CE}, T_s)$ computes the set of suspicious statements $\f{SuspStats}$ whose mutation may lead to generate a valid patch. 
It finds these suspicious statements by combining the counterexamples \f{CE} generated by termination prover \f{TP} with the results of
spectrum-based fault localization~\cite{jones2002:visualization,naish2011:model,heiden2019:evaluation,xie2021:essential} on the test suite
(executed with a timeout mechanism to avoid getting stuck in infinite loops);
(ii) the function $\f{mutate}(P_s, \f{SuspStats})$ is used to construct the patch space (i.e., patch generation) by mutating the computed set $\f{SuspStats}$ that may affect the truth value of the termination condition of the detected buggy non-terminating  loop in $P_s$.

\begin{figure}[b]
\vspace*{-2ex}\small
\algorule\par
\begin{algorithmic}[1]
\State \textbf{Inputs}: $P_s,~ \varphi_{beh}, ~ T_s,~ \f{TP},~ \f{TimeBudget}$ 
\State \textbf{Output}: $P_\f{repaired}$
\State $\f{PatchSpace} := \emptyset,~ P_\f{repaired} := \f{NoPatchFound},~ \f{found} := \f{false}$
\State $(*, \f{CE}) := \f{TP}(P_s)$                     \Comment{get CE, verdict known to be \f{NT}}
\State $\f{SuspStats} := \f{faultLocalizer} (P_s, \f{CE}, T_s) $  %
\State $\f{PatchSpace} := \f{mutate}(P_s, \f{SuspStats})$          \Comment{patch generation}
\While{$\f{PatchSpace} \neq \emptyset \land \f{TimeBudget} > 0 \land \neg \f{found} $}
\State \textbf{select} $\f{patch}$ \textbf{from} $\f{PatchSpace}$
\If{$\f{patch} \models \varphi_{beh}$}                              \Comment{2-step patch validation}
\If{$\f{TP} (\f{patch}) = (\f{TR}, *)$}
\State $P_\f{repaired} := \f{patch}$
\State $\f{found} := \f{true}$
\EndIf
\EndIf
\EndWhile
\State \Return $P_\f{repaired}$
\end{algorithmic}
\vspace*{-1ex}
\algorule\vspace*{-1ex}
\caption{Repair algorithm for sequential programs}
\label{alg:RepairAlgSeq}
\end{figure}

\subsection{Termination Bugs in Concurrent Programs}\label{termConcur}

\noindent
In the automated repair of concurrent programs, the goal is to generate a patch that ensures that a concurrent program is correct under all interleavings. It is difficult, if not impossible, to examine all possible executions of a concurrent program using dynamic analysis techniques. Therefore, a concurrent program cannot be debugged and repaired in the same manner as sequential programs. In the case of concurrency, it usually refers to action interleaving. That is, if the processes $p_i$ and $p_j$ are in parallel composition ($p_i \parallel p_j$) then the actions of these will be interleaved.
Each process executes a sequence of actions (sub-program), then the set of possible interleavings of several processes consists of all possible sequences of actions. 
Before proceeding further, let us introduce the notion of successful termination in concurrent programs.

\begin{definition}[Termination of Concurrent Programs]
Let $P_c = (p_1 \parallel p_2 \parallel ... \parallel p_n)$ be a concurrent program that consists of a collection of sub-programs, where each process $p_i$ executes a sub-program. 
Let $H_{i}$ be the set of halting statements at the sub-program executed by process $p_i$.
We say that  $P_c$ is terminating iff every sub-program is eventually terminating.
That is, the sub-program executed by process $p_i$  eventually executes some halting statement $s \in H_{i}$ and terminates.
\end{definition}

\begin{definition}[Termination Failures in Concurrent Programs]
Let $P_c = (p_1 \parallel p_2 \parallel ... \parallel p_n)$ be a concurrent program that consists of a collection of sub-programs executed by processes $p_1,..., p_n$.
The program $P_c$ fails to terminate iff:
\begin{compactenum}
\item there exists a logical bug that leads to an infinite loop, or
\item there exists a concurrency bug, such as deadlock or livelock, that prevents the program from making any further progress to reach a halting statement and terminate.
\end{compactenum}
\end{definition}
Thus, if deadlocks and livelocks are formally proven to never occur in the program-under-analysis, 
and all loops in the sub-programs are proven to be terminating, 
then one can conclude that the concurrent program is terminating.
The distinction between the two causes of termination failures in concurrent programs (logical bug or concurrency bug) helps to select a strategy for fixing the detected termination bug.

\head{Combining Model Checking and Termination Provers}
Repairing termination bugs in concurrent programs can be a computationally complex task.
This is mainly because termination bugs in concurrent programs can be caused by either a logical or concurrency bug. 
Furthermore, the vast number of possible interleavings of parallel processes of a given concurrent program can increase the complexity of the repair problem.
Therefore, it is necessary to employ both termination provers and model checking to reduce the computational complexity of the problem.
Fortunately, we know how to write specifications to check the absence of concurrency bugs in concurrent programs~\cite{gupta2018:model,lin2014:automatic, zhou2017:undead}. 
In Definition \ref{ValidityConcurrentSystems}, we describe the conditions that are necessary to ensure the correctness of the generated patches for termination bugs in the buggy non-terminating concurrent program.

\begin{definition}[Validity of Generated Patches for Termination Bugs in Concurrent Programs]\label{ValidityConcurrentSystems}
Let $P_c = (p_1 \parallel p_2 \parallel ... \parallel p_n)$ be a buggy non-terminating concurrent program, $\f{TP}$ be a termination prover,
and $\f{SMC}$ be a software model checker.
Let $\varphi_\f{deadlock}$ and $\varphi_\f{livelock}$ be specifications that check respectively the absence of deadlocks and livelocks in $P_c$ and $\varphi_\f{beh}$ be a specification that captures the expected behavior of $P_c$.
We say that a candidate patch $P'_c$ for the buggy program $P_c$ is a valid patch iff it meets the following requirements: 
\begin{compactenum}
\item the prover $\f{TP}$ returns ``terminating'' when analysing termination of the sub-program executed by process $p_i$, and
\item the  checker $\f{SMC}$ returns ``holds'' when checking the specifications $\varphi_\f{deadlock}$ and $\varphi_\f{livelock}$ against $P'_c$, and
\item the checker $\f{SMC}$ returns ``holds''  when checking the specification of expected behavior $\varphi_\f{beh}$ against the patch $P'_c$.
\end{compactenum}
\end{definition}

Formally, we can capture the above described requirements in a 4-part correctness specification of the following form:
\begin{equation}\label{SpecConcur}
\begin{array}[t]{l}
\varphi_\f{con} = 
\forall_{p_i \in P'_c} (\f{TP}(p_i) = \f{TR}) \land \f{SMC}(P'_c , \varphi_\f{deadlock}) ~ \land \\ \hspace{41pt} \f{SMC}(P'_c , \varphi_\f{livelock}) \land \f{SMC}(P'_c , \varphi_\f{beh})
\end{array}
\end{equation}
A key challenge when dealing with termination bugs is to ensure that the generated repair guarantees not only the termination of the program for each possible input, but also the semantic preservation of the program.
This requires the analysis of a composite correctness property that checks both termination and semantic preservation.
By using formula (\ref{SpecConcur}),
we entirely avoid the patch overfitting problem, which is one of the major problems of dynamic APR.
Formula (\ref{SpecConcur}) uses sequential termination provers to check the absence of infinite loops in each individual process. 
This can be performed while abstracting away concurrency details that are irrelevant to the local computations of processes. 
Note that one cannot prove the termination of program $P_c$ by simply applying a sequential termination prover: 
a sound proof of termination must consider all possible interactions among the sub-programs of $P_c$.

An alternative way to use model checking in detecting and repairing termination bugs of the concurrent program is to reduce the termination problem to the reachability analysis problem. 
That is, to check whether each process will eventually reach some halting location and terminate.
However, the feasibility of the approach relies mainly on the size and number of processes of the buggy program under analysis 
(i.e., computing state-reachability is known to be PSPACE-complete when processes are finite state~\cite{kozen1977:lower}).
The reduction of the termination problem to the reachability problem in model checking leads to the following temporal formula
\begin{equation} \label{concReach}
\varphi_\f{con}' = \forall_{p_i \in P'_c} (\textbf{AF} (p_i.\ell_h^{(j)} \mid \ell_h^{(j)} \in H_i))\land \f{SMC}(P'_c , \varphi_\f{beh})
\end{equation}
where $\textbf{A}$ is a temporal path quantifier which should be read as ``for all paths'', 
and $\textbf{F}$ is the ``future'' temporal operator~\cite{pnueli1977:temporal}. 
Intuitively, formula (\ref{concReach}) checks whether for each reachable execution path of  process $p_i$,
some halting location $\ell_{h}^{(j)} \in H_i$ will eventually be reached. 
However, the proper termination analysis of the concurrent program $P_c$ using formula (\ref{concReach}) requires the precise computation of the set of halting locations $H_i$ for each process $p_i$ of the patched concurrent program $P'_c$.

There are several software model checkers\footnote{~Unlike traditional model checking, 
a software model checker does not require a user to manually construct an abstract model of the program to be checked, 
but instead, the tool works directly on the program's source code.} 
that can be used to verify reachability properties and detect concurrency bugs, including VeriSoft~\cite{godefroid1997:verisoft}, 
Java Pathfinder~\cite{havelund2000:model}, CMC~\cite{musuvathi2002:cmc}, DIVINE~\cite{baranova2017:model}, and GMC~\cite{thompson2008:verification}. 
DIVINE is a modern, explicit-state model checker that can verify programs written in multiple real-world programming languages, including C and C++.
On the other hand, GMC is a model checker based on the generic Monte-Carlo model-checking algorithm. 
It takes as input a C program, the target program to be verified, and the linear temporal logic specification that needs to be checked.

There are also a few termination provers that can be used to analyze the termination of concurrent programs. 
For instance, Cook et al.~\cite{cook2007:proving} have extended the termination prover T2~\cite{brockschmidt2016:t2} to support the analysis of concurrent programs, 
which can be used to validate generated patches for termination bugs in concurrent programs. 
Termination checker T2 supports nested loops, recursive functions, pointers, side-effects, and function-pointers, as well as concurrent programs. 
Of course, the prover cannot handle termination of all concurrent programs since the general problem is undecidable.
The use of concurrent termination provers leads to the following specification
\begin{equation} \label{concurTP}
\varphi_\f{con}'' = \forall_{p_i \in P'_{c}} (\f{TP} (p_i) = \f{TR}) \land \f{SMC}(P'_c , \varphi_\f{beh})
\end{equation}
The termination provers AProVE, 2LS, and T2 can be viewed as complementary tools: it is possible that some tool fails to detect certain forms of termination bugs while other succeeds, depending on the implemented theory and the complexity of the program under analysis. Therefore, termination checkers can be run in parallel to expose termination bugs.

While checking the satisfaction of formula (\ref{SpecConcur}) may require higher computational complexity than formulas (\ref{concReach}) and (\ref{concurTP}) (i.e., it employs both termination provers and software model checkers to check the absence of deadlocks, livelocks, and infinite loops), it has several advantages.
First, it helps identify the root causes of non-termination in the program-under-repair. 
Second, the patch validation approach that uses formula (\ref{SpecConcur}) can benefit from the counterexamples generated by both termination provers and software model checkers. 
This helps to develop effective program synthesis for termination bugs.

\head{Repairing Termination Bugs in Concurrent Programs}
To fix termination bugs in concurrent programs, it is first essential to identify the root cause of the termination bug since both logical and concurrency bugs can cause non-termination.
On the one hand, there are several mechanisms for handling deadlocks in concurrent programs~\cite{lin2014:automatic,zhou2017:undead,cai2016:fixing}.
On the other, repair algorithms based on genetic programming can help fix termination bugs that occur due to infinite loops~\cite{legoues2012:genprog,yu2020:smart}. 

Figure~\ref{alg:RepairAlgConcur} sketches a hybrid repair procedure for termination bugs in concurrent programs. 
It uses three helper functions:
(i) $\f{faultLocalizer}(P_c, \f{CE})$ finds the statements that are suspected to be faulty in $P_c$ using the counterexamples \f{CE} generated by termination prover \f{TP} and software model checker \f{SMC},
(ii) $\f{mutateConcur}(P_c, \f{SuspStatsD})$ constructs the patch space for a concurrency bug by mutating the synchronization primitives of the program, and 
(iii) $\f{mutateLogic}(P_c, \f{SuspStatsL})$ constructs the patch space for a logical bug by mutating  expressions that affect the control of faulty loops.

\begin{table*}
\centering
\caption{A comparison of the complexity of the repair problem of termination bugs in sequential and concurrent programs} \label{table:comparisonTable}
\vspace*{-2ex}%
\resizebox{\linewidth}{!}{%
\begin{tabular}{ccccc} 
\toprule
\textbf{program class} & \textbf{ root causes of non-termination} & \textbf{feasible APR} & \textbf{patch validation procedure} & \textbf{ patch validation tools} \\
\midrule 
sequential programs & infinite loops & dynamic-static & test cases and termination provers & AProVE, 2LS, T2, and GMC \\
concurrent programs & infinite loops, deadlocks, or livelocks & static or formal & termination provers and SMC & T2 and GMC \\
\bottomrule
\end{tabular}}%
\vspace*{-0ex}%
\end{table*}

\begin{figure}[ht]
\vspace*{.5ex}\algorule\small
\begin{algorithmic}[1]
\State \textbf{Inputs}: $P_c,~ \varphi_\f{beh},~ \varphi_\f{deadlock},~ \f{TP},~ \f{SMC},~ \f{TimeBudget}$ 
\State \textbf{Output}: $P_\f{repaired}$
\State $\f{PatchSpace} := \emptyset,~ P_\f{repaired} := \f{NoPatchFound}$,~ \f{found} := \f{false}
\State $(\f{Outcome}_1, \f{CE}_1) := \f{SMC}(P_c,  \varphi_\f{deadlock})$
\If{$ ( \f{Outcome}_1 = \f{fails}$)}
\Comment{$P_c$ contains a deadlock}
\State $\f{SuspStatsD} := \f{faultLocalizer} (P_s, \f{CE}_1) $
\State $\f{PatchSpace} := \f{mutateConcur}(P_c, SuspStatsD)$
\EndIf
\State $(\f{Outcome}_2, \f{CE}_2) := \f{TP}(P_c)$
\If{$\f{Outcome}_2 = \f{NT}$}                 \Comment{$P_c$ contains an infinite loop}
\State $\f{SuspStatsL} := \f{faultLocalizer} (P_s, \f{CE}_2) $
\State $\f{PatchSpace} := \f{PatchSpace} \cup \f{mutateLogic}(P_c, \f{SuspStatsL})$
\EndIf
\While{$\f{PatchSpace} \neq \emptyset \land \f{TimeBudget} > 0 \land \neg \f{found} $}
\State \textbf{select} $\f{patch}$ \textbf{from} $\f{PatchSpace}$
\If{$\f{patch} \models \varphi_\f{beh}$}
\Comment{2-step patch validation}
\If{$\f{TP}(\f{patch}) = (\f{TR},*) \land \newline
\hspace*{1.25cm}\f{SMC}(\f{patch},  \varphi_\f{deadlock}) = (\f{holds},*)$}
\State $P_\f{repaired} := \f{patch}$,~ $\f{found} := \f{true}$
\EndIf
\EndIf
\EndWhile
\State \Return $P_\f{repaired}$
\end{algorithmic}%
\vspace*{-1ex}\algorule\vspace*{-1ex}%
\caption{Repair algorithm for concurrent programs}%
\label{alg:RepairAlgConcur}%
\vspace*{-1ex}%
\end{figure}

Generating verified repairs of termination bugs in both sequential and concurrent programs is a challenging open problem that requires formal analysis techniques.
Our preliminary investigation of using termination provers in sequential programs shows that this is a rather promising approach that can be effectively integrated with dynamic APR approaches to generate verified repairs for termination bugs. 
The application of the termination provers 2LS and AProVE on the programs in the two datasets, SNU real-time benchmark and the Power-Stone benchmark suite~\cite{ku2007:buffer}, show that the tools are able to successfully prove termination of around $85\%$ of the examined programs using very little computational time (a few seconds).
This demonstrates the feasibility of using termination provers to validate the generated patches of termination bugs.

Table \ref{table:comparisonTable} compares the complexity of termination bugs in both sequential and concurrent programs using several criteria: 
(i)  root causes of termination bugs,
(ii) feasible APR approach to be applied for termination bugs in both classes of programs, 
(iii) patch validation procedure to validate generated patches,
(iv) patch validation tool that can be used to check termination. 

\section{Related Work}

\noindent
We discuss the related literature on APR, bug classification,  use of formal specifications in APR,  previous attempts to integrate different analysis techniques, and termination analysis.

\head{Automated Program Repair}
Source-based, automated program repair approaches~\cite{monperrus2018:automatic} can be separated into search-based and semantic-based approaches. 
\emph{Search-based} approaches such as GenProg~\cite{legoues2012:genprog}, Astor~\cite{martinez2016:astor}, 
and SCRepair~\cite{yu2020:smart} predominantly use failing test cases to identify bugs, 
and then mutate the source code until the program passes all failing test cases.
They do not provide patch correctness guarantees beyond the fact that the provided test cases now pass. 
Recent work introduced property-based testing to strengthen the validation of candidate repairs and address overfitting~\cite{gissurarson2022:propr}.
Nevertheless, these approaches require executing the program-under-repair, first to find the bug, and then to generate and validate candidate repairs. 
\emph{Semantic-based} approaches like SemFix~\cite{nguyen2013:semfix}, Nopol~\cite{xuan2017:nopol}, DirectFix~\cite{mechtaev2015:directfix}, SPR~\cite{long2015:staged}, Angelix~\cite{mechtaev2016:angelix}, and JFIX~\cite{le2017:jfix} infer repair constraints for the buggy program via symbolic execution of the given tests. 
The completeness of inferred constraints relies on the available test suite.
Similarly, Infinitel~\cite{marcote2015:automatic} uses an SMT solver to synthesise a loop termination condition and then uses test cases for patch validation.

\head{Bug Classification Systems} 
Several works target bug classification using a wide variety of classification criteria and for different goals.
The work of Tan et al.~\cite{li2006:have, tan2014:bug} introduced a bug classification system based on the cause-impact criteria.
They studied bug characteristics of around 2,060 real-world bugs in three representative open-source projects.
They concluded that semantic bugs are the dominant root cause of bugs, and memory-related bugs have decreased due to the development of effective detection tools. 
Many bug tracking systems classify bugs using severity and priority criteria, where severity indicates the seriousness of the bug on the program functionality and priority indicates how soon the bug should be fixed~\cite{serrano2005:bugzilla}.
Cotroneo et al. present a maintenance-oriented bug classification system in which the characteristics of the bug manifestation are studied~\cite{cotroneo2016:how}. 
The study identifies the set of failure-exposing conditions under which a bug may occur. 
Neither of these classification systems considers properties that can be used to analyze the detection power of different bug detection techniques and the conditions under which they can be integrated, which we add with the classification system proposed in this paper.
Abbaspour et al.~\cite{asadollah2015:classification} do use bug observability to study the erroneous behavior of a wide variety of concurrency bugs. 
However, the authors restrict themselves to concurrency bugs and do not study the detection power of different bug detection techniques.

\head{Integrating Bug Detection Techniques}
Few attempts have been made to integrate different program analysis techniques to alleviate the impact of the patch overfitting problem.
Al-Bataineh et al. used the static detection patterns/rules as the source for formulating formal specifications and discussed the possibility of integrating static and dynamic analysis techniques to improve the overall quality of generated patches~\cite{al-bataineh2021:reliable}. 
There also exists a few examples of using information from debugging to aid APR:
Facebook's APR tool SapFix takes information generated during the bug detection process and applies various techniques, including templates specific to given bug types, to fix the program~\cite{marginean2019:sapfix}. 

\head{Termination Analysis of Programs}
A huge body of work has been published on proving termination of programs based on a variety of techniques, 
such as abstract interpretation~\cite{berdine2007:variance, chawdhary2008:ranking, tsitovich2011:loop}, 
bounds analysis~\cite{gulwani2009:controlflow, gulwani2009:speed},
ranking functions~\cite{bradley2005:linear, cousot2005:proving}, 
recurrence sets~\cite{gupta2008:proving, harris2010:alternation}, 
and transition invariants~\cite{kroening2010:termination, podelski2004:transition}. 
Based on these techniques, a number of program termination checkers have been developed in the prior literature including AProVE~\cite{giesl2014:proving}, 2LS~\cite{chen2015:synthesising}, T2~\cite{brockschmidt2016:t2}, and ARMC~\cite{podelski2007:armc}. 
Unfortunately, termination provers have not yet been used to \emph{validate} the generated candidate patches of termination bugs in the previous APR approaches.
We strongly believe that integrating APR approaches with contemporary termination checkers would help advance the current state-of-the-art of APR, 
not only for repairing termination bugs, but also for other classes of concurrency bugs.
This is mainly because fixing termination bugs in concurrent programs would ensure the absence of certain classes of concurrency bugs, such as deadlock and livelock bugs.
It would also help avoid the patch overfitting problem by generating verified repairs for termination bugs in both sequential and concurrent programs. 

\section{Concluding Remarks}

\noindent
A significant class of bugs cannot be handled with current APR approaches, 
and there is a need to study complementary techniques.
To stimulate this work, 
we propose a novel bug classification system based on three key properties: 
bug observability, bug tractability, and bug reproducibility.
This provides a tool to methodologically
explore and compare alternative and hybrid APR approaches by 
(i) analyzing the detection power of different bug detection techniques, 
(ii) distinguishing APR approaches based on their bug repair capabilities, and 
(iii) providing a common terminology that helps identify gaps in current APR research.
Moreover, it allows analysis of how techniques can be combined to handle challenging classes of bugs. 
As a demonstrating example, we study termination bugs in sequential and concurrent programs, 
and present novel \emph{hybrid} algorithms for their repair by integrating termination provers and software model checkers in the APR pipeline.
Such integration reduces the complexity of the repair algorithms 
and improves the overall reliability.

We identify the following directions for future research:
(1) we are in the process of empirically validating the ideas described in this work 
by combining a selection of APR tools with some of the termination provers and software model checkers mentioned earlier;
(2) there is a need for efficient fault localization mechanisms for termination and liveness bugs, 
since these create infinite traces that cannot be handled with the current spectrum-based fault localization approaches;
We currently circumvent this with a timeout mechanism, but more efficient techniques could find a more precise set of suspicious statements;
(3) the combination of CounterExample-Guided Inductive Synthesis (CEGIS)~\cite{solar-lezama2006:combinatorial} with termination provers and software model checkers could enable efficient patch space exploration without the user guidance that is normally needed for CEGIS;
(4) aside from the termination bugs studied in Section~\ref{sec:HybridTermination}, 
various other bugs cannot be handled by APR approaches solely based on dynamic analysis. 
One particularly interesting class of non-observable bugs for future research are security-related vulnerabilities 
where sensitive information may be disclosed to unauthorized parties as a result of violations of information flow security~\cite{sabelfeld2003:languagebased,smith2007:principles}.

{\balance
\printbibliography
}
 
\end{document}